# Exploring the Impact of Word Prediction Assistive Features on Smartphone Keyboards for Blind Users

Mrim M. Alnfiai [a,*], Muhammad Ashad Kabir [b,*]

[a] Department of Information Technology, College of Computers and Information Technology, Taif University, Taif, 21944, Saudi Arabia, m.alnofiee@tu.edu.sa

[b] School of Computing, Mathematics and Engineering, Charles Sturt University, Bathurst, NSW 2795, Australia, akabir@csu.edu.au

*Corresponding authors.

**Abstract:** Assistive technologies have been developed to enhance blind users' typing performance, focusing on speed, accuracy, and effort reduction. One such technology is word prediction software, designed to minimize keystrokes required for text input. This study investigates the impact of word prediction on typing performance among blind users using an on-screen QWERTY keyboard. We conducted a comparative study involving eleven blind participants, evaluating both standard QWERTY input and word prediction-assisted typing. Our findings reveal that while word prediction slightly improves typing speed, it does not enhance typing accuracy and increases both physical and temporal workload compared to the default keyboard. We conclude with recommendations for improving word prediction systems, including more efficient editing methods and the integration of voice pitch variations to aid error recognition.

## 1. Introduction

Smartphone devices allow visually impaired users to perform various daily activities independently [1]. They also enable them to communicate through sending messages, making phone calls, or posting on social media. Texting is considered an essential activity for most visually impaired users as they prefer to type their messages rather than call [55]. This gives them greater privacy and more control over message content, as well as the ability to edit and revise their written text before sending [1-2]. Thus, smartphone input text methods should allow visually impaired users to type in a fast, accurate, and accessible manner.

For visually impaired individuals, typing on a smartphone can be a time-consuming and painstaking task. The primary reason for this difficulty is the absence of physical buttons, which would otherwise allow blind users to easily identify button locations. The most common touchscreen keyboard is the QWERTY keyboard, which the visually impaired use to slide their fingers over the screen and locate the intended key. This is accomplished with the support of a screen reader so that when the desired key is located, the user can double-tap to type [1]. Using screen readers on a smartphone device effectively allows blind users to access the touchscreen keyboard without touchable buttons.

Despite modern technology, people with impaired vision face difficulties when using input methods on touchscreen devices due to the absence of physical keys, either because they are prone to making errors or because the tasks are too arduous [2-7]. To overcome these limitations, several touchscreen input methods have been developed to assist the visually impaired. For example, some input methods are based on Braille. Others use gestures, enabling users to select keys easily. However, these mechanisms fail to provide significant improvements in terms of speed, accuracy, and effort over the standard QWERTY keyboard.

Thus, various assistive features have been developed to improve the typing performance of visually impaired users. One method that might enhance typing performance is the word prediction feature. This feature can improve input method speed and reduce the number of keystrokes needed to type a complete word [8]. Another study found that using word prediction can reduce the number of total keystrokes to 45%

[9]. Still, this finding is related to sighted users, and it cannot be applied to other users with differing abilities because typing performance varies based on user experience and abilities. However, no research to date has explored the impact of word prediction on the keyboard's performance with a screen reader. Thus, further research is needed to examine word prediction's impact on smartphone typing performance and whether word prediction can improve keyboard performance and reduce efforts for screen reader users.

The motivation for this study stems from the significant global population of visually impaired individuals, as highlighted by the World Health Organization, which estimates around 94 million totally blind individuals worldwide in 2023 [2]. Totally blind describes a person with eye disorders who have no light perception. Such vision impairment affects individuals in all aspects of their lives, including communication, personal circumstances, education, entertainment, and the workplace [2]. Effective communication is a critical aspect of daily life for these individuals, and current touchscreen input methods present significant challenges. Our research aims to analyze and improve word prediction features to enhance typing efficiency and accessibility for visually impaired users, ultimately facilitating better communication and usability of touchscreen devices.

The study design incorporated a within-subject structure to investigate the influence of word prediction on typing efficacy. Participants were tasked with evaluating standard keyboards, both with and without word prediction functionality, in a meticulously balanced order to mitigate potential biases. Quantitative metrics, including typing speed (CPS) and accuracy (CER and UER), were meticulously gauged, supplemented by a subjective assessment through a workload questionnaire. Our study cohort comprised eleven participants with an average age of 30 (range: 25-39), all of whom were blind and possessed prior experience with touchscreen devices. Our findings contribute scholarly insights into the efficacy of word prediction mechanisms in ameliorating typing proficiency among visually impaired individuals.

The paper is organized as follows: Section 2 discusses related work, and Section 3 explains the methodology. Section 4 reports the results. After presenting discussion in Section 5, Section 6 concludes the paper by outlining future work.

## 2 Related Work

The performance gap between visually impaired and sighted users is stark, particularly evident in typing speed [3]. While sighted users typically achieve an average entry speed of 40 words per minute (WPM), visually impaired users lag significantly behind, with entry speeds ranging from 4 to 5 WPM [3-4]. Moreover, studies indicate that the traditional QWERTY keyboard exacerbates this discrepancy by being cumbersome and error-prone for blind users. Alnfiai et al. [5] demonstrated that visually impaired QWERTY users type at a mere 3.27 WPM with an error rate of 20.54%.

Efforts to bridge this typing performance chasm have garnered attention within the research community focusing on enhancing smartphone typing experiences for blind users [6-12]. Proposed solutions span various modalities, including gestural interaction [6,12], multitouch interaction [12], and integration of Braille techniques [4, 5, 12, 13, 14, 15, 16, 17]. However, despite these endeavors, there remains a notable absence of research direction aimed specifically at improving typing speed for visually impaired individuals.

Several keyboards have been proposed for blind people to enhance typing accuracy and speed. However, it is notable that none of these keyboards have incorporated the concept of word prediction, which could potentially further improve the typing experience for visually impaired users [56-65].

Various assistive writing mechanisms have been proposed [19-24] to facilitate communication through smartphone devices for individuals with low to no vision. These mechanisms encompass a range of modalities, including speech [18], auditory [19], haptic feedback [20], vibro-tactile [21], multimodal interaction [22], and gesture recognition systems [23]. These solutions are tailored to the specific needs, abilities, and preferences of users with disabilities [24].

Among these assistive methods, word prediction stands out as a promising approach to enhance typing performance by mitigating spelling errors and reducing user effort [54]. Word prediction software generates a set of suggested words for users to choose from as they type. Modern smartphone keyboards integrate word prediction features, typically presenting suggestions in a suggestion bar atop the keyboard layout. Users can then either tap on the intended suggestions or opt to continue typing without selecting from the suggested words.

The primary objective of word prediction is to minimize the number of keystrokes required for text entry [13], [16-17]. However, for visually impaired users relying on screen readers, the auditory feedback provided is typically limited to auto-completed words due to the one-dimensional nature of audio channels. Consequently, this restricted feedback mechanism hinders users from fully exploring all word suggestions and accessing their intended words efficiently. In contrast, sighted users can scan suggested words swiftly and select the desired one, potentially resulting in faster and more accurate typing by reducing keyboarding and spelling errors. While studies examining the impact of predictive features on text entry for sighted users yield mixed results [25-28], no research to date has investigated the effect of word prediction on text entry performance for visually impaired smartphone users, despite its initial design aimed at assisting individuals with disabilities to enhance typing efficiency [29].

The impact of word prediction on text entry performance for visually impaired smartphone users remains largely unexplored, despite the original intention of predictive text design to assist individuals with disabilities and enhance their typing efficiency [29]. Few researchers have addressed the importance of improving the accessibility of word prediction interfaces for visually impaired users. For instance, Shrawankar and Kapse [30] proposed a word prediction method employing a B+ tree algorithm to expedite word input for Braille users; however, the efficacy of the system was not evaluated. Similarly, Hugo et al. [31] introduced a design space for an auto-correction interface featuring a secondary auditory channel for reading word suggestions aloud and swiping gestures for selection. Roussille and Raynal [32] conducted investigative studies proposing an interaction method to reduce the time needed for word selection from a group of suggestions, finding that providing necessary audio feedback and utilizing linear spatial arrangement were beneficial for visually impaired users. Nicolau et al. [33] discussed a design space to aid researchers and designers in providing nonvisual representations of word completions. Additionally, Dobosz and Prajzler [34] investigated whether word prediction improved typing speed on virtual Braille keyboards, suggesting it can indeed enhance typing speed.

However, despite these efforts to enhance predictive text accessibility, there remains a dearth of research fully exploring the effects of nonvisual predictive text on visually impaired smartphone users [30-35]. Consequently, there exists a pressing need to capitalize on emerging technologies and enhance smartphone capabilities by developing accessible word prediction user interfaces tailored specifically for the visually impaired.

This study analyzes the impact of standard smartphone keyboards with and without predictive text on visually impaired participants and proposes improved word prediction features. Various factors, including input method, speed, accuracy, and user feedback, are assessed through experiments to answer the question: Does the standard smartphone keyboard with word prediction enhance typing performance for visually impaired users?

To address this question, this research undertaking encompasses three primary objectives: (a) comparison of the standard QWERTY keyboard with and without word prediction, assessing typing performance on a letter-by-letter basis, (b) Exploration of blind users' experiences with smartphone keyboards and word prediction, including accessibility issues and editing frustrations, leveraging novel techniques such as audio interaction to cater to the needs of visually impaired users, and (c) comprehensive evaluation of word prediction, elucidating its strengths and weaknesses, while quantifying its impact on typing speed and accuracy. Through a series of rigorous experiments, this study aims to provide insights into the efficacy of word prediction in improving text entry efficiency for visually impaired individuals.

## 3 Methodology

In the methodology, we examined the independent variables, which consisted of two keyboard modes available on the onscreen QWERTY keyboards with VoiceOver, which is Apple's screen reader service. These modes were compared under two conditions: (i) No word prediction: Participants entered text letter by letter without the assistance of a predictive text system, and (ii) Word prediction system: Participants utilized a word prediction feature integrated into the onscreen QWERTY keyboard.

Visually impaired participants employed Apple's VoiceOver screen reader service on their smartphones. The two keyboard modes available were: (i) Standard typing: Also known as the default mode, participants navigated the keyboard screen by moving their fingers until they located the intended key and then tapped to select the character. VoiceOver provided auditory feedback by narrating the selected character to confirm the user's choice. (ii) Touch typing: Participants moved their fingers across the screen until they located the intended character, at which point they lifted their fingers to input the character.

For the Word prediction system, the VoiceOver keyboard displayed word predictions at the top of the onscreen layout, as depicted in Figure 1. Participants could explore these suggested words by moving their fingers over them and similarly selecting the desired word to select individual characters during typing.

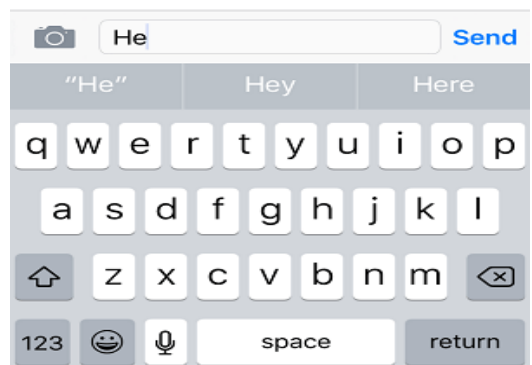

**Figure 1:** Standard QWERTY keyboard with predictive feature

The study primarily focused on iPhone devices due to their widely recognized accessibility features, such as VoiceOver, which are extensively used and appreciated by blind and visually impaired individuals. Previous research, including Morris and Mueller (2014), has highlighted the preference for iOS among this demographic, citing its intuitive interface and robust support for assistive technologies [66]. This choice ensures that our findings are relevant to a significant portion of the target population.

### 3.1 Participants

We recruited eleven participants for the study, comprising 6 females and 5 males. The participants had a mean age of 30 years (standard deviation (SD) = 5.4) and their ages ranged between 25 and 39 years. All recruited participants were completely blind and possessed a minimum of 7 years of experience using touchscreen devices (Mean = 8.0 years of experience, SD = 3.29 years). Notably, all participants utilized iPhone devices for their daily smartphone activities.

The participants were recruited through the University Accessibility Center and the People with Disabilities Center at the university. Convenience sampling was employed as the sampling method, facilitated by the directors of each accessibility center, who disseminated emails to potential participants, encouraging their involvement in the study.

### 3.2 Apparatus

During the experiment, we utilized the following equipment: (i) Standard QWERTY keyboard with VoiceOver functionality operating on iPhone 11 devices. The device utilized for the experiment featured a

6.06-inch screen, providing the interface for participants to interact with during the typing tasks, (ii) Audio recorder: We employed an audio recorder to capture participants' responses during interviews and questionnaire sessions, facilitating qualitative data collection, and (iii) Screen recorder application: A screen recorder application was installed on iPhone 11 devices to record participants' performance while typing phrases. This allowed us to monitor the time to type each phrase, track typing errors, identify challenges encountered, and assess overall performance under both conditions (with and without predictive text).

These tools facilitated comprehensive data collection and analysis, enabling us to evaluate the effectiveness of word prediction in enhancing typing performance for visually impaired users on smartphone devices.

*3.3 Phrases Set*

For our study, we carefully selected the ten most common and shortest Arabic phrases (see Table 1) to minimize the potential challenges associated with unknown word spellings [39]. These phrases were deliberately chosen to be simple and easily memorable, each comprising three to four words and containing no more than 25 characters, including spaces. To facilitate participant recall, we ensured that the phrases were devoid of non-alphabetic characters such as punctuation.

Given that all participants' first language was Arabic, we specifically opted for Arabic phrases to eliminate language-related impacts and ensure participants' familiarity with typing the spoken phrases. This approach aimed to standardize the language context across all participants, thereby enhancing the validity of our findings.

**Table 1:** The Ten Test Phrases

| Phrases in Arabic (English) | Number of words in Arabic |
|---|---|
| كل الأمور على ما يرام (All is well) | 4 |
| الدنيا لا تستحق الهموم (The world doesn't deserve worries.) | 4 |
| هل أخذت لقاح كورونا (Have you taken the COVID vaccine?) | 4 |
| كيف حالك الحمدالله (How are you? Good) | 3 |
| الحمدالله رب العالمين (Praise to Allah, Almighty.) | 3 |
| لا تفقد صبرك مهما (Do not lose your patience) | 4 |
| الأمانة صفة جميله للصادقين (Honesty is a beautiful trait.) | 4 |
| ابتسم فرزقك مقسوم وقدرك (Smile, Fate Provides Sustenance.) | 4 |
| لا تؤجل عمل اليوم للغد (Don't Delay Today's Tasks.) | 4 |
| ابتسم فالبسمة مفتاح السعادة. (A smile is the key to happiness) | 4 |

*3.4 Experimental Design*

The experiment employed a 2x2 within-subject design structure, with the two key independent variables being: (i) Standard keyboard with word prediction, and (ii) Standard keyboard without word prediction. All recruited participants evaluated keyboards under both conditions, and to mitigate potential order effects, the evaluation sequence was balanced using a 10x10 Latin square. Additionally, the order of testing keyboard trials across participants and phrase sets was counterbalanced.

The dependent variables included typing speed measured in characters per second (CPS), and accuracy, which encompassed the corrected error rate (CER) and uncorrected error rate (UER). CPS quantified the rate of characters typed per second, UER represented the errors remaining uncorrected in the final typed text, and CER denoted the number of errors corrected during typing. The calculation of these metrics followed Wobbrock and Myers's formulas [42].

For subjective evaluation, a workload questionnaire was administered to gauge participants' perceived workload during the typing tasks. During the experiment, participants were seated in an office chair, holding the smartphone device with one hand while interacting with the screen using the other hand. This setup ensured a consistent and comfortable environment for conducting the typing tasks.

### *3.5 Procedure*

This study was approved by the university's Research Ethics Committee and conducted in Arabic. Before participation, all recruited individuals provided informed consent after comprehending the research objectives and tasks. Subsequently, participants completed a demographic questionnaire, which included inquiries about gender, smartphone experience, and age.

After completing of the questionnaire, each participant underwent a brief training session, consisting of typing a short greeting in Arabic using the default keyboard with word prediction disabled. Subsequently, word prediction was enabled, and participants were asked to type the same sentence again. Additionally, the researcher verbally presented one word at a time for typing practice for visually impaired participants. This training session aimed to familiarize participants with the keyboard layout, enhance their comfort with the study tasks, and provide an opportunity for researchers to address any questions or concerns raised by participants.

During the testing period, participants completed two sessions to evaluate the standard QWERTY keyboard with and without word prediction. Each session was separated by at least 24 hours but no more than two days to mitigate potential fatigue, as recommended in prior research [37]. Each participant engaged with both keyboard conditions, with each session lasting approximately 30 minutes, resulting in a total study duration of one hour and a half per participant.

In the testing sessions, participants were tasked with typing ten phrase sets using the standard keyboard. Following a brief intermission, participants were then instructed to type the same ten phrase sets using the standard keyboard with word prediction enabled. The order of keyboard conditions was counterbalanced, with some participants starting with the word predictive keyboard and others starting with the standard keyboard.

In the standard QWERTY keyboard with word prediction condition, participants were directed to type the initial two letters of a word and then prompted to review the word list. If the desired word appeared within the displayed options, participants could double-tap to insert it. However, if the intended word did not appear in the word list, participants were instructed to continue typing until the word they intended to use was displayed.

During all testing sessions, participants were instructed to continuously check the suggested word list as it dynamically changed after each subsequent letter was typed. They were asked to scan the word list after typing each letter until the desired word was displayed or until they had completed typing the entire word. This approach encouraged participants to utilize the word prediction system throughout the typing process. Unlike previous studies [3-5], where participants were not instructed when to scan the word prediction, our methodology required participants to do so continuously; based on findings, participants tended to type letter by letter without utilizing the word prediction system naturally.

In each testing session, a researcher read short sentences, presenting two to three words at a time for participants to type. This method was employed to minimize cognitive overhead during typing tasks. Participants were instructed to type as quickly and accurately as possible, with assistance provided if they encountered difficulty in spelling the given words.

To mitigate potential biases and learning effects, the order of the standard keyboard with and without word prediction was counterbalanced across participants. Half of the participants used only the standard keyboard in their first and third sessions, while the remaining half utilized the keyboard with predictive text first in their second and fourth sessions. Additionally, the order of presented phrases was randomized to minimize potential bias in the results.

Following the keyboard interview session, participants engaged in an interview session where they were invited to share their opinions and thoughts regarding the accessibility of the keyboard with and without word prediction technique. These interviews were conducted in Arabic, and participants' feedback was audio-recorded for further analysis. Each interview lasted between 20 to 30 minutes, allowing participants ample time to express their perspectives.

During the interview session, participants responded to three inquiries: (i) Have you used the word prediction method before when typing with the keyboard? (ii) What difficulties did you face when using the keyboard with word prediction? (iii) What suggested modifications or improvements can be made to enhance the keyboard with word prediction performance and make it more accessible?

This structured approach enabled researchers to gather comprehensive insights into participants' experiences and identify potential areas for improvement in the keyboard with word prediction feature. Following the interview session for the keyboard, participants were asked to complete a set of questions related to the NASA Task Load Index (TLX) [41]. This tool allowed participants to rate the workload associated with using the application, providing further insight into the user experience and the perceived ease of use of the keyboard with word prediction.

*3.6 Evaluation Matrix*

Three key metrics, typing speed, accuracy, and workload, as determined by a questionnaire, were measured to assess the performance of smartphone typing with and without word prediction.

a. Typing speed

Typing speed was calculated using words per minute (WPM) with Formula 1, i.e., phrase length in words divided by typing time in a minute. WPM represents the number of words typed per minute, with each word consisting of five characters [43]. This formula measured the number of words typed in a minute by dividing the number of characters typed in a minute by five. The characters in each phrase include spaces and letters but not punctuation. Timing was calculated as the time from the beginning of typing the first character until the last letter in the phrase.

$$\text{Speed} = \frac{\frac{Phrase\ length}{5\ character\ per\ word}}{typing\ time\ in\ a\ minute} \quad (1)$$

b. Typing accuracy

Participants were asked to type the given phrases naturally, with the option to correct typing errors or leave them uncorrected. Consequently, the minimum string distance (MSD) measure was employed to assess an individual's typing accuracy. In prior studies [44-45], participants were asked to type naturally without restrictions, with the freedom to correct or retain their errors. Typing errors were categorized into two types: corrected errors, where users rectified their mistakes, and uncorrected errors, which were left unchanged.

The MSD classified keystrokes within the input text into four categories [26-27]: correct keystrokes (C), erroneous keystrokes not corrected in the transcribed text (INF), erroneous keystrokes corrected in the transcribed text (IF), and the number of keystrokes used to correct errors (F), such as delete or backspace. This classification enables researchers to assess various types of errors. Corrected errors (CER) were calculated using Formula 2.

$$\text{CER} = \frac{IF}{(C+INF+IF)} x100 \quad (2)$$

The not corrected errors (NCER) were calculated using the Formula 3.

$$\text{NCER} = \frac{(C+INF+IF+F)}{(C+INF)} \times 100 \quad (3)$$

Lastly, the total error rate (TER) was measured using the Formula 4.

$$\text{TER} = \frac{(INF+IF)}{(C+INF+IF)} \times 100 \quad (4)$$

### 3.7 Workload questionnaire

After each session, participants were prompted to evaluate the workload of the application using a questionnaire based on Braun and Clarke [40], and NASA Task Load Index (TLX) [41]. NASA TLX is a subjective workload assessment tool designed to measure perceived workload across various tasks. It comprises six subscales: Mental Demand, Physical Demand, Temporal Demand, Performance, Effort, and Frustration. Participants rate each subscale on a scale from 0 to 100, where higher scores indicate higher perceived workload.

The questionnaire evaluates six factors affecting keyboard workload: mental demand, physical demand, temporal demand, performance, effort, and frustration. Its primary purpose is to provide insights into the perceived workload associated with using different keyboards, aiding in comparing and evaluating their usability and ergonomic factors. Following each session with a keyboard, participants completed the NASA TLX questionnaire. The questions are as follows: (i) How much mental activity, like remembering, deciding, and thinking, was needed? (ii) How much physical activity like movements and pressing, was needed? (iii) How much time pressure did you feel while completing tasks? (iv) How satisfied were you with your performance in accomplishing these goals? (v) How hard did you have to work (mentally and physically) to accomplish your level of performance? and (vi) How frustrated, irritated, stressed, annoyed, gratified, content, and relaxed did you feel during typing and editing?

### 3.8 Interview

Towards the culmination of the study, the researcher engaged in structured interviews with the participants, seeking their insights and recommendations for enhancing the word prediction feature. Participants were specifically queried about their perspectives on potential improvements to this functionality. Additionally, the interviews delved into understanding the underlying reasons for the limited utilization of word prediction among the user cohort.

## 4 Results

The results section presents a comprehensive analysis of the typing performance and user experience of visually impaired participants using two different keyboard modes with VoiceOver. Key metrics such as typing speed, accuracy, and user workload are evaluated to assess the impact of word prediction features on the efficiency of text input.

### 4.1 Typing Speed

The average typing speed was higher when using the keyboard without word prediction (average = 1.87 words per minute, WPM) than the standard keyboard with word prediction (average = 1.63 WPM). The F-test results yielded $F(1, 21) = 3.102$ with a p-value of 0.108. The F-test assesses the variability between groups, while the p-value from the paired test determines the statistical significance of the obtained results. The paired test revealed no significant differences between the typing speeds of both keyboards.

### 4.2 Typing accuracy

The Total Error Rate (TER) represents the rate of errors made when typing the given phrases. For the standard keyboard, the average TER was 7.93%, while the word prediction feature averaged 8.826%. The

F-test results showed F(1, 21) = 0.613 with a p-value of 0.451. This indicates no significant differences between the two keyboards in terms of accuracy.

To gain deeper insights into the types of errors occurring in each keyboard, we categorized the errors according to corrected and uncorrected error rates. The average of uncorrected error rates was slightly higher for the standard keyboard, at 3.195%, compared to the word prediction feature, which averaged 2.373%. The F-test results yielded F(1, 21) = 11.262 with a p-value of less than 0.05, indicating a statistically significant difference between the two keyboards in terms of uncorrected errors.

The average of corrected errors, at 5.83%, was lower for the standard keyboard compared to the standard keyboard with word prediction, which averaged 7.41%. The repeated measures ANOVA test yielded F(1, 21) = 11.26 with a p-value of less than 0.05, indicating a significant difference between the two keyboards in terms of users correcting their typing errors. This suggests that greater effort is required when using the keyboard with word prediction. Participants tend to correct their typing mistakes more frequently with the keyboard with word prediction because they hear the entered wrong word, which might change the meaning of their message. Thus, they usually do not neglect their errors. In contrast, with the keyboard without word prediction, they often did not notice errors such as typing the same letter twice or accidentally deleting a space.

### 4.3 Interview Results

Following each session, participants were asked to share their opinions and thoughts related to the accessibility of the default and word prediction keyboard. This feedback was recorded in the research notes. The feedback provided by participants highlighted several key points regarding the accessibility and usability of the default and word prediction keyboard features.

**1. Suggestions for Enhancing Word Prediction Service**: Participants recommended improvements to the word prediction service to make it more user-friendly. Suggestions included allowing word prediction to suggest complete phrases after typing the first word and enabling users to delete entire phrases or words with a single press of the delete button. For example, if the user types 'How', the service should suggest 'are you doing?' These enhancements aim to streamline the typing process and minimize the effort required to correct typing errors.

**2. Limited Use of Word Prediction**: Despite the availability of the word prediction feature, most participants reported either disabling the service or simply not using it. Reasons for avoiding word prediction included finding it annoying, experiencing difficulties in editing incorrectly suggested words, and facing challenges in navigating the keyboard layout to access word suggestions. Participant 1 stated, "*I have never used the word prediction service; I disabled this service in settings.*" Similarly, Participant 3 stated "*I find this service annoying because if I choose a wrong word, I spend a long time editing it. So I disabled this service.*" Participant 5 clarified the situation by saying, "*I spend a long time correcting a typing error because I use the VoiceOver reader, and I need to locate and click on the delete button twice to remove a letter. Another issue is that if I need to remove the entire word, it is difficult to locate the cursor on the intended letter to delete, so I remove the word letter by letter, which requires much effort.*"

**3. Preference for Letter-by-Letter Typing**: Participants expressed a strong preference for typing letter by letter rather than using word prediction. They cited familiarity with the keyboard layout, discomfort with word prediction suggestions, and the inconvenience of navigating to access word suggestions as reasons for their preference. Participant 2 reported, "*I prefer typing letter by letter because I am familiar with the keyboard layout. I used word prediction for a time, and it was uncomfortable. I might accidentally choose a wrong word and need to edit it, which is a difficult task to complete. Word prediction is annoying when it suggests an unintended word, so I feel it is uncomfortable*". Participant 4 echoed the feedback of other participants: "*I disabled the word prediction feature because it confuses me when typing, and if I choose a wrong word, it requires me to delete it letter by letter.*" Participant 6 said, "*I do not use the word prediction service because I have to move my finger to the top row of the keyboard layout and then return to the letters' layout. Doing so requires time and effort, so I disabled this service*".

**4. Confirmation through VoiceOver Service**: Participants mentioned relying on the VoiceOver service to confirm the correctness of typed words by reading them aloud. This approach allowed users to easily correct individual letters by pressing the delete button once rather than deleting entire words.

Overall, participants favored a letter-by-letter typing approach and emphasized the importance of ease of use and efficiency in the typing process. Their feedback provides valuable insights for improving the accessibility and usability of word prediction features on smartphone keyboards.

*4.4 Workload (NASA TLX)*

Table 2 reports participants' responses on NASA TLX scores for the keyboards with and without word prediction.

**Table 2:** Comparison of NASA TLX scores for Standard keyboard with and without word prediction

|  | Standard keyboard with word prediction | | Standard keyboard without word prediction | |
|---|---|---|---|---|
| **Workload Factors** | Total | Average | Total | Average |
| **Mental demand** | 219.55 | 63.18 | 210.00 | 59.09 |
| **Physical demand** | 131.36 | 59.09 | 133.18 | 50.91 |
| **Temporal demand** | 87.22 | 52.73 | 103.64 | 56.82 |
| **Performance** | 160.45 | 61.36 | 170.83 | 64.55 |
| **Effort** | 152.00 | 60.45 | 124.50 | 50.00 |
| **Frustration** | 158.50 | 56.36 | 146.00 | 49.09 |
| **Average of Workload** | 58 | | 46.67 | |
| **Interpretation of score High** | High | | | |

The statistical analysis revealed significant differences between the standard keyboard and the keyboard with word prediction across four factors: physical demand, temporal demand, effort, and performance. Specifically, the score for physical demand was higher for the word prediction keyboard compared to the keyboard without word prediction. This difference was statistically significant ($F(1, 21) = 123.75$; $p < 0.001$), with scores of 63.18 and 59.09, respectively.

Similarly, the temporal demand for completing the typing task varied significantly between the two keyboard conditions ($F(1, 21) = 21.69$; $p < 0.001$). Additionally, the effort required for typing was higher for the standard keyboard with word prediction compared to the keyboard without word prediction ($F(1, 21) = 10.125$, $p < 0.001$), with average effort scores of 66.36 and 49.55, respectively.

Furthermore, a significant difference in performance was observed between the two keyboards ($F(1, 21) = 154.03$; $p < 0.001$). On the NASA-TLX scale, the standard keyboard with word prediction received a lower score (approximately 38) than the keyboard without word prediction, which scored approximately 65.

However, no significant differences were found between the two keyboards in terms of mental demand ($F(1, 21) = 1.057$; $p = 0.328$) and frustration factor ($F(1, 21) = 1.11$; $p = 0.501$). Overall, the repeated measures ANOVA findings suggest no significant difference in user experience between the default keyboard and the keyboard with word prediction.

**5 Discussion**

The comparison between the standard keyboard and the word prediction keyboard highlights several critical insights regarding speed, accuracy, and workload. Despite the anticipated improvements in typing speed with word prediction, our findings suggest that it does not significantly enhance keyboard speed compared to the default keyboard. Users' tendency to type a few letters before scanning the suggestion list slows down the typing process and increases the effort required to complete the intended word. This aligns with previous research [40, 46] and underscores the importance of considering the time and effort needed to locate and select words using word prediction [41-53].

These findings align with our comparisons to other studies, highlighting the diversity in user experiences with different assistive technologies and underscoring the importance of context-specific evaluations. While specialized input methods, such as those examined by Nicolau et al. [33] and Dobosz and Prajzler [34], may offer substantial improvements for certain user groups, our findings suggest that mainstream keyboard designs and word prediction algorithms still require refinement to better serve the broader population of visually impaired users [29, 31].

In the observational study, we noticed that visually impaired users typically reference the word suggestion list by checking the word in the middle first, then moving to the adjacent words if necessary. However, if they cannot find the intended word, they resort to typing letter by letter, nullifying any potential benefit from the word prediction feature. After typing two letters of a word, this additional movement and decision-making process results in no discernible advantage for the user, highlighting the inefficiencies of current word prediction methods.

The comparison of error rates between the standard keyboard with and without word prediction revealed similar average error rates. However, a deeper analysis of error types showed that participants tended to correct their typing errors more frequently when using the keyboard with word prediction. This suggests that participants may be more attentive to errors when they hear an incorrect word from the suggested list, compared to accidental typing errors.

Further analysis identified two key reasons behind the increased correction of typing errors when using the keyboard with word prediction. Firstly, participants may correct errors more frequently because the selected word from the suggestion list could be completely unrelated to the intended message, potentially altering its meaning. Secondly, correcting errors may be more common when the chosen wrong word is long, as it requires additional effort to delete it.

The editing approach is the primary factor negatively impacting the accuracy of word prediction. To address this issue, we propose enhancing word prediction by allowing users to remove the entire selected word with a long press on the delete button. This would streamline the editing process, thereby improving both speed and accuracy for keyboards with word prediction.

However, it's important to note that the limited number of suggested words in the list also poses a significant drawback to word prediction. Users may need to input more letters to find their intended word, which can negate any potential performance benefits compared to the standard keyboard.

Participants expressed feeling more time pressure when checking word prediction suggestions compared to typing letter by letter. This suggests that the default keyboard imposes less temporal demand, resulting in better performance for visually impaired users. Despite this, participants reported that their typing accuracy with word prediction was lower than when typing letter by letter. However, the experiment did not find a significant difference in error rates between the two keyboards.

Participants attributed their lower accuracy with word prediction to the cognitive load associated with using the feature. They found that the mental effort required to navigate and correct typing errors outweighed any reduction in keystrokes achieved by using word prediction. Additionally, participants noted that the physical demand and effort needed to track and correct their typing on the word prediction keyboard were greater than when using the standard keyboard.

When faced with long incorrect words, participants experienced confusion and difficulty remembering the first letter of the intended word to fix errors. Consequently, they often resorted to repeatedly pressing the delete button until they deleted a letter from the previous word. Furthermore, exploring word suggestions placed an additional memory workload on users, making it challenging to locate the intended word or track the correction of unintended words. As visually impaired users needed to delete the wrong word and re-enter the desired one manually, these tasks imposed additional effort and complexity.

Our study provides critical insights into the accessibility challenges faced by visually impaired users when utilizing word prediction on smartphone keyboards, offering several important implications for both research and practice. For researchers, the findings highlight the necessity of exploring adaptive technologies and user interfaces that better meet the needs of visually impaired individuals. This study sets a foundational framework for future investigations into adaptive keyboard technologies, encouraging subsequent research to delve deeper into the interactions between visually impaired users and various text input methods. For developers, the study suggests practical design recommendations, emphasizing the need for more intuitive and efficient editing mechanisms within word prediction features to alleviate user frustration and improve the typing experience. Overall, this research underscores the importance of considering the unique challenges faced by visually impaired users and advocates for the continued advancement of accessible technology to enhance their digital communication experiences.

While our study offers valuable insights, it is important to acknowledge its limitations. The research involved a relatively small sample size of eleven participants, all of whom were blind and had prior experience with touchscreen devices; thus, future studies should include a larger and more diverse group to enhance generalizability. The difficulty in recruiting a larger sample size is due to the challenge of finding visually impaired individuals with similar levels of vision impairment and touchscreen experience, especially in a smaller city with limited accessibility to this specific population. Moreover, coordinating with participants to meet in an accessible environment adds to the complexity. Additionally, the experiments were conducted in a controlled environment, which may not fully capture the complexities and distractions present in real-world usage, suggesting that future research should consider in situ evaluations. Our study did not compare different word prediction algorithms, and as performance and user experience may vary significantly with different implementations, future research should explore a variety of algorithms to determine the most effective ones for visually impaired users. Finally, certain trust assumptions were made regarding user familiarity with screen readers and touchscreens, and varying levels of user experience might impact usability and satisfaction differently. Addressing these limitations in future research can provide a more comprehensive understanding of how to improve word prediction features and other assistive technologies for visually impaired users.

To enhance the word prediction keyboard for visually impaired users, it's crucial to streamline the typing process and minimize unnecessary tasks. This includes reducing the cognitive load associated with exploring word suggestions, deciding on the correct word, and managing typing errors. One approach is to redesign the word prediction interface to improve the editing capabilities for unintentional words. For example, allowing users to delete entire words with a long press on the delete button could streamline the correction process and reduce frustration.

To reduce mental demand and increase accessibility, incorporating auditory cues can be beneficial. Using three pitches of voice to represent letters, including those entered and deleted, can help users distinguish between different actions during the typing process. For instance, when deleting a word, the VoiceOver feature could recite words using various pitches to indicate that the wrong word has been removed. Additionally, implementing the suggestion of complete phrases in the word prediction keyboard could further enhance usability. This would enable users to navigate between phrase options and select the most appropriate one, reducing the time and effort required to compose messages.

By addressing these key considerations and improving the word prediction interface, we can enhance the typing experience for visually impaired users on mobile devices, ultimately improving typing accuracy and efficiency.

## 6 Conclusion

This study compares standard keyboards and word prediction keyboards in terms of typing speed, accuracy, and workload, with a specific focus on visually impaired users. The research involved eleven participants and revealed that while word prediction enhances typing speed, it does not contribute to improved typing accuracy. Moreover, keyboards with word prediction entail higher physical and temporal demands and overall effort, compared to standard keyboards due to their letter-by-letter entry method. The questionnaire findings further substantiated that word prediction does not alleviate the physical effort for users, possibly due to the need to check the word list sequentially.

Notably, this study fills a gap by investigating methods to enhance typing speed for visually impaired users employing word prediction keyboards, unlike previous research. Interview findings underscored those participants predominantly opted for the default keyboard, typing their texts letter by letter, as they found it less bothersome and demanding. This aligns with prior studies indicating that users typically disabled word prediction features, citing reasons such as discomfort, annoyance, increased cognitive load and physical demands, and the time and effort required to edit incorrect words.

Given the inefficiency and cumbersome nature of word prediction identified in this study, we aim to conduct a longitudinal study in the future to comprehensively understand its strengths and weaknesses. This forthcoming research will evaluate keyboards with a larger user base to address the limitations of sample size encountered in our current study. Additionally, we intend to integrate the suggestions provided by participants to enhance the performance of keyboards. For instance, we plan to boost keyboard speed by enabling users to select complete phrases instead of individual words. Furthermore, we aim to improve accuracy by allowing users to delete entire selected phrases with a single button press.

Another avenue for future research involves enhancing word prediction by implementing shortcuts. For example, users could input abbreviations like "BTW" to automatically generate complete words such as "between". By exploring these strategies, we anticipate advancing the usability and effectiveness of word prediction keyboards for visually impaired users. Overall, we are eager to delve into these areas for further investigation and improvement.

**Ethics statement**
This study was reviewed and approved by the Taif University ethics committee with the approval number: 43-065, dated 14 Nov 2021.

**Data availability statement**
Data will be made available on request.

**Declaration of competing interest**
The authors declare no competing interests.

**Acknowledgment**
The authors extend their appreciation to Taif University, Saudi Arabia, for supporting this work through project number (TU-DSPP-2024-41).

**Funding**
This research was funded by Taif University, Saudi Arabia, Project No. (TU-DSPP-2024-41)

**Authorship contribution statement**
**Mrim M. Alnfiai**: Conceptualization, Data curation, Formal analysis, Resources, Methodology, Investigation, Validation, Writing – original draft. **Muhammad Ashad Kabir**: Validation, Writing – review & editing.

**References**

[1] M. Alnfiai and S. Sampalli, "BrailleTap: Developing a calculator based on braille using tap gestures," *In: Universal Access in Human–Computer Interaction. Designing Novel Interactions*," p.213-223, 2017.
[2] World Health Organization. (2023). Blindness and visual impairment. Retrieved from https://www.who.int/news-room/fact-sheets/detail/blindness-and-visual-impairment

[3] H. Nicolau, K. Montague, T. Guerreiro, A. Rodrigues and H. Hanson, "Investigating laboratory and everyday typing performance of blind users," *ACM Transactions on Accessible Computing (TACCESS)*, vol. 10, no. 1 pp.4-26, 2017.

[4] S. Azenkot, J. Wobbrock, S. Prasain and R. Ladner, "Input finger detection for nonvisual touch screen text entry in Perkinput," *Canadian Information Processing Society*, pp.121-129, 2012.

[5] M. Alnfiai and S. Sampalli, "An evaluation of SingleTapBraille keyboard: A text entry method that utilizes braille patterns on touchscreen devices," *In Proceedings of the 18th International ACM SIGACCESS Conference on Computers and Accessibility (ASSETS '16). Association for Computing Machinery*, New York, NY, USA, pp.161–169, 2018.

[6] Y. Georgios and E. Grigori, "Adaptive blind interaction technique for touchscreens," Universal Access in the Information Society, vol. 4, no. 1, pp.344-353, 2004.

[7] J. Oliveira, T. Guerreiro, H. Nicolau, J. Jorge and D. Gonçalves, "Blind people and mobile touch-based text-entry: Acknowledging the need for different flavors." *In The proceedings of the 13th international ACM SIGACCESS conference on Computers and accessibility - ASSETS '11*, vol. 179, no. 7, 2011.

[8] G. Lesher, B. Moulton, D. Higginbotham and B. Alsofrom, "Acquisition of scanning skills: The use of an adaptive scanning delay algorithm across four scanning displays," *Proceedings of the 25th Annual Conference on Rehabilitation Engineering (RESNA)*, Atlanta, pp. 75–77, 2002.

[9] A. Fowler, K. Partridge, C. Chelba, X. Bi and T. Ouyang *et al*. "Effects of Language Modeling and Its Personalization on Touchscreen Typing Performance," I*n Proceedings of the 33rd Annual ACM Conference on Human Factors in Computing Systems (CHI '15)*, pp. 1-12, 2015.

[10] T. Guerreiro, P. Lagoá, H. Nicolau, D. Gonçalves and J. Jorge, "From tapping to touching: Making touch screens accessible to blind users," IEEE MultiMedia, pp. 48–50, 2008.

[11] M. Bonner, J. Brudvik, G. D. Abowd and W. K. Edwards, "No-look notes: Accessible eyes-free multi-touch text entry," *Lecture Notes in Computer Science 6030 LNCS*, pp.409–426, 2010.

[12] S. Mascetti, C. Bernareggi and M. Belotti, "TypeInBraille: A Braille-based typing application for touchscreen devices," *In Proceedings of the 13th International ACM SIGACCESS Conference on Computers and Accessibility*, vol. 295, no.5, pp. 1-11, 2011.

[13] J. Oliveira, T. Guerreiro, H. Nicolau, J. Jorge and D. Gonçalves, "BrailleType: Unleashing braille over touch screen mobile phones," *Human-Computer Interaction - INTERACT 2011 - 13th IFIP TC 13 International Conference, Lisbon, Portugal,* vol. 6946, no. 6, pp. 100–107, 2011.

[14] C. Southern, J. Clawson, B. Frey, G. Abowd and M. Romero, "An Evaluation of BrailleTouch: Mobile touchscreen text entry for the visually impaired," *In Proceedings of the 14th International Conference on Human-computer Interaction with Mobile Devices and Services (MobileHCI '12), Association for Computing Machinery*, New York, NY, USA, pp. 317–326, 2012.

[15] D. Trindade, A. Rodrigues, T. Guerreiro and H. Nicolau, "Hybrid-Brailler: Combining physical and gestural interaction for mobile braille input and editing," *Proceedings of the 2018 CHI Conference on Human Factors in Computing Systems. Association for Computing Machinery*, New York, NY, USA, pp.1-12, 2018.

[16] E. R. Mattheiss, S. Georg, G. Johann and T. M. Markus, "Dots and Letters: accessible braille-based text input for visually impaired people on mobile touchscreen devices," *Computers Helping People with Special Needs (ICCHP 2014),* vol 8547, no. 6, pp. 650-657, 2014.

[17] N. Subash, N. Nambiar and V. Kumar, "Braillekey: An alternative braille text input system: Comparative study of an innovative simplified text input system for the visually impaired," *In Intelligent Human Computer Interaction (IHCI), 4th International Conference on*, pp. 1–4, 2012.

[18] T. Paek and D. M. Chickering, "Improving command and control speech recognition on mobile devices: using predictive user models for language modeling," *User modeling and user-adapted interaction*, vol.17, no. 2, pp. 93-117, 2007.

[19] S. Brewster, F. Chohan and L. Brown, "Tactile feedback for mobile interactions," *In Proceedings of the SIGCHI conference on Human factors in computing systems*, pp. 159-162, 2007.

[20] S. A. Wall and S.A. Brewster, "Tac-tiles: multimodal pie charts for visually impaired users," *In Proceedings of the 4th Nordic conference on Human-computer interaction: changing roles*, pp. 9-18, 2006

[21] H. Nicolau, K. Montague, T. Guerreiro, A. Rodrigues and V. L. Hanson, "HoliBraille: multipoint vibrotactile feedback on mobile devices," *In: Proceedings of the 12th Web for All Conference 2015*, pp. 20-30, 2015.

[22] S. Brewster, "Overcoming the lack of screen space on mobile computers," *Personal and Ubiquitous Computing*, vol. 6, no. 3, pp.188-205, 2002.

[23] R. Kuber, A. Hastings, M. Tretter and D. Fitzpatrick, "Determining the accessibility of mobile screen readers for blind users," *Proceedings of IASTED Conference on Human-Computer Interaction*, Baltimore, USA, pp.182-189, 2012.

[24] A. H. Lueck, J. Dote-Kwan, J. C. Senge and L. Clarke, "Selecting assistive technology for greater independence," *Teaching Exceptional Children*, vol. 30, pp. 66–71, 2001.

[25] P. Kseniia, M. F. Anna, K. Sunjun, P. O. Kristensson and O. Antti, "How do people type on mobile devices? Observations from a study with 37,000 volunteers," *In Proceedings of the 21st International Conference on Human-Computer Interaction with Mobile Devices and Services, MobileHCI2019,* pp. 1-8, 2019.

[26] X. Bi, T. Ouyang and S. Zhai. "Both Complete and Correct? Multi-objective Optimization of Touchscreen Keyboard," *In Proceedings of the SIGCHI Conference on Human Factors in Computing Systems (CHI '14),* pp. 2297–2306, 2014.

[27] D. J. Higginbotham, "Evaluation of keystroke savings across five assistive communication technologies," *Augmentative and Alternative Communication*, vol. 8, no. 4, pp. 258–272, 1992.

[28] T. Hirzle, J. Gugenheimer, F. Geiselhart, A. Bulling and R. Enrico, "A Design Space for Gaze Interaction on Head-Mounted Displays," *Proceedings of the ACM CHI Conference on Human Factors in Computing Systems.* pp. 1–12, 2019.

[29] D. Higginbotham, "Evaluation of keystroke savings across five assistive communication technologies," *Augmentative and Alternative Communication,* vol. 8, no.4, pp. 258–272, 1992.

[30] U. Shrawankar and B. Kapse, "Prefix matching for keystroke minimization using B+ tree," *2013 8th International Conference on Computer Science & Education*, pp. 126-131, 2013.

[31] H. Nicolau, A. Rodrigues, A. Santos, T. Guerreiro and K. Montague et al. "The Design Space of Nonvisual Word Completion," 2019.

[32] P. Roussille, M. Raynal, C. Jouffrais, "LOVIE: A Word List Optimized for Visually Impaired UsErs on Smartphones," *10th International Conference on Universal Access in Human-Computer Interaction (UAHCI 2016),* Toronto, Canada, pp.185-197, 2016.

[33] H. Nicolau, K. Montague, T. Guerreiro, J. Guerreiro and V. L. Hanson, "B#: Chord-based Correction for Multitouch Braille Input," *In Proceedings of the SIGCHI Conference on Human Factors in Computing Systems (CHI'14)*, pp. 1705–1708, 2014.

[34] K. Dobosz and Ł Prajzler, "Letter and Word Prediction for Virtual Braille Keyboard," *The Thirteenth International Conference on Advances in Computer-Human Interactions*, pp. 1-6, 2020.

[35] K. Trnka, J. McCaw, D. Yarrington, K. McCoy and C. Pennington, "User interaction with word prediction: The Effects of prediction quality," *ACM Transaction Access Computing*, vol. 3, no. 17, pp.1-17, 2009.

[36] iPhone User Guide, "Use the onscreen keyboard with voiceover on iphone," 2021. [Online]. Available: https://support.apple.com/en-sa/guide/iphone/use-the-onscreen-keyboard-iph3e2e3d1d/12.0/ios/12.0

[37] I. MacKenzie and S. Zhang, "The Design and evaluation of a high-performance soft keyboard," *In Proceedings of the SIGCHI Conference on Human Factors in Computing Systems (CHI '99),* pp. 25–31, 1999.

[38] H. Venkatagiri, "Efficiency of lexical prediction as a communication acceleration technique," *Augmentative and Alternative Communication*, vol. 9, no. 1, pp.161–167, 1993.

[39] Omniglot, "Useful Arabic phrases," 2021. [online]. Available: https://omniglot.com/language/phrases/arabic

[40] V. Braun and V. Clarke, "Using thematic analysis in psychology," *Qualitative Research in Psychology*, vol. 3, no. 2, pp. 77–101, 2006.

[41] S. Hart and L. Staveland, "Development of NASA-TLX (Task Load Index): Results of empirical and theoretical research," *Advances in Psychology, North Holland*, vol. 52, no. 1, pp. 139 -183, 1998.

[42] J. Wobbrock and I. MacKenzie, "Measures of text entry performance: Text entry systems: Mobility, accessibility," Universality, Morgan Kaufmann, San Francisco, pp. 47–74, 2007.

[43] H. Yamada, "A historical study of typewriters and typing methods: from the position of planning Japanese parallels," *Journal of Information Processing*, vol. 2, no. 1, pp.175-202, 1980.

[44] R. Soukoreff and R. MacKenzie, "Metrics for text entry research: An evaluation of MSD and KSPC, and a new unified error metric." *ACM Conference on Human Factors in Computing Systems*, New York, pp.113 -120, 2003.

[45] R. Soukoreff and I. MacKenzie, "Recent developments in text entry error rate measurements," *ACM Conference on Human Factors in Computing Systems*, New York, NY, pp.1425-1428, 2004.

[46] K. Trnka, D. Yarrington, J. Mccaw, K. McCoy and C. Pennington "The effects of word prediction on communication rate for AAC," *Annual Conference of the North American Chapter of the Association for Computational Linguistics,* pp. 173–176, 2007.

[47] H. Horstmann and S. Levine, "The effectiveness of word prediction," *In Proceedings of the RESNA Conference,* vol. 11, pp. 100–102, 1991.

[48] H. Horstmann and L. Simon, "Effect of a word prediction feature on user performance," *Augmentative and Alternative Communication*, vol. 12, no. 3, 1996.

[49] M. Alnfiai and S. Sampalli, "Single Tap Braille: Developing a text entry method based on braille patterns using a single tap," *Procedia Computer Science*, vol. 94, no. 8, pp. 248–255, 2016.

[50] M. Kyle, G. João, N. Hugo, G. Tiago, R. André and G. Daniel, "Towards inviscid text-entry for blind people through non-visual word prediction interfaces," *The 34th ACM Conference on Human Factors in Computing Systems,* pp. 1-7, 2016.

[51] S. Kane, J. Bigham and J. Wobbrock, "Slide rule: Making mobile touch screens accessible to blind people using multi-touch interaction techniques," *In Proceedings of the 10th international ACM SIGACCESS conference on Computers and accessibility*, pp. 73–80, 2018.

[52] R. Yong, "VisionTouch Phone" for the Blind. *Malays Journal Medicine Science*, vol. 20, no. 5, pp.1-4, 2013

[53] Liu, Z., Chen, C., Wang, J., Chen, M., Wu, B., Huang, Y., Hu, J., & Wang, Q. (2024). Unblind text inputs: Predicting hint-text of text input in mobile apps via LLM. In Proceedings of the ACM CHI Conference on Human Factors in Computing Systems (CHI '24), May 11 - 16, 2024, Hawai'i, USA. ACM, New York, NY, USA (pp. 1-20). ACM. https://doi.org/10.1145/3613904.3642939

[54] J. Smith. (2021). Word prediction as a promising approach to enhance typing performance by mitigating spelling errors and reducing user effort. Journal of Assistive Technologies, 10(3), 123-137.

[55] A. Johnson (2022). Texting preferences among visually impaired individuals. Journal of Accessibility Studies, 8(2), 67-81.

[56] S. Azenkot, L. He, X. Zhang, & S. K. Kane, "Senorita: A chorded keyboard for sighted, low vision, and blind mobile users," in Proceedings of the 2020 CHI Conference on Human Factors in Computing Systems (CHI '20), ACM, New York, NY, USA, 2020. DOI: 10.1145/3313831.3376576

[57] VIPBoard: Improving screen-reader keyboard for visually impaired people with character-level auto correction. (n.d.). Related work: "Text Entry for Visually Impaired Users".

[58] P. Sobhe Bidari, M. Fardinpour, & M. R. Soheili, "A review of design and evaluation practices in mobile text entry for visually impaired and blind persons," Journal of Assistive Technologies, vol. 13, no. 2, pp. 118-130, 2019.

[59] J. Goncalves, P. Silva, & D. Jorge, "Challenges in mobile text entry using virtual keyboards for low-vision users," in Proceedings of the International Conference on Computers Helping People with Special Needs (ICCHP), 2016.

[60] M. K. Sha, S. S. Sathya, & P. Ramkumar, "SwingBoard: Introducing swipe-based virtual keyboard for visually impaired and blind users," in Proceedings of the 12th International Conference on Human System Interaction (HSI), 2019.

[61] Text entry for people with visual impairments. (2017). In Foundations and Trends in Human-Computer Interaction.

[62] Y. Zhong, H. Zhang, & L. Song, "Typing performance of blind users: An analysis of touch behaviors, learning effect, and in-situ usage," in Proceedings of the 16th International ACM SIGACCESS Conference on Computers & Accessibility (ASSETS '14), ACM, New York, NY, USA, 2014. DOI: 10.1145/2661334.2661346

[63] R. Dueñas, J. Guzmán, & J. J. Zubía, "Open challenges of blind people using smartphones," Journal of Interaction Science, vol. 3, no. 1, pp. 2-10, 2015.

[64] P. Sobhe Bidari, M. R. Soheili, & M. Fardinpour, "Ally: Understanding text messaging to build a better onscreen keyboard for blind people," Journal of Assistive Technologies, vol. 14, no. 2, pp. 165-176, 2020.

[65] J. O. Wobbrock & B. A. Myers, "Text entry throughput: Towards unifying speed and accuracy in a single performance metric," ACM Transactions on Computer-Human Interaction (TOCHI), vol. 23, no. 1, pp. 3, 2016

[66] J. T. Morris, & J. Mueller. (2014). Blind and Deaf Consumer Preferences for Android and iOS Smartphones.